\documentclass[lettersize,journal]{IEEEtran}
\usepackage{algorithmic}
\usepackage{algorithm}
\usepackage{array}
\usepackage{textcomp}
\usepackage{stfloats}
\usepackage{url}
\usepackage{verbatim}
\usepackage{cite}

\usepackage{amsmath,amssymb,amsfonts}
\usepackage{algorithmic}
\usepackage{graphicx}
\usepackage{textcomp}
\usepackage{soul}
\usepackage{caption}
\usepackage{subcaption}
\usepackage{xcolor}
\usepackage{comment}
\usepackage{tabularx}
\usepackage{array}
\usepackage{multirow}
\usepackage{float}
\usepackage{multicol}

\graphicspath{ {./Figures/} }

\begin{document}

\title{Implementation of a Modified U-Net for Medical Image Segmentation on Edge Devices}
\author{Owais Ali, Hazrat Ali,~\IEEEmembership{Senior Member,~IEEE}, Syed Ayaz Ali Shah, Aamir Shahzad
\thanks{Owais Ali, Hazrat Ali, Syed Ayaz Ali Shah, Aamir Shahzad are with Department of Electrical and Computer Engineering, COMSATS University Islamabad, Abbottabad Campus, Abbottabad Pakistan. 
}
\thanks{Correspondence to: Hazrat Ali. Email: hazratali@cuiatd.edu.pk}
\thanks{Final version available from IEEE. Use doi: 10.1109/TCSII.2022.3181132. IEEE Copyrights apply}
\thanks{Manuscript received January 12, 2022; revised xxxx xx, 2022.}}

\markboth{IEEE Transactions on Circuits and Systems II: Express Briefs, ~Vol.~xx, No.~xx, August~2021}%
{Owais \MakeLowercase{\textit{et al.}}: Edge Computing for Medical Image Segmentation}


\maketitle

\begin{abstract}
Deep learning techniques, particularly convolutional neural networks, have shown great potential in computer vision and medical imaging applications. However, deep learning models are computationally demanding as they require enormous computational power and specialized processing hardware for model training.   
To make these models portable and compatible for prototyping, their implementation on low-power devices is imperative. In this work, we present the implementation of Modified U-Net on Intel/Movidius Neural Compute Stick 2 (NCS-2) for the segmentation of medical images. We selected U-Net because, in medical image segmentation, U-Net is a prominent model that provides improved performance for medical image segmentation even if the dataset size is small. The modified U-Net model is evaluated for performance in terms of dice score. Experiments are reported for segmentation task on three medical imaging datasets: BraTs dataset of brain MRI, heart MRI dataset, and Ziehl–Neelsen sputum smear microscopy image (ZNSDB) dataset. For the proposed model, we reduced the number of parameters from 30 million in the U-Net model to 0.49 million in the proposed architecture. Experimental results show that the modified U-Net provides comparable performance while requiring significantly lower resources and provides inference on the NCS-2. The maximum dice scores recorded are 0.96 for the BraTs dataset, 0.94 for the heart MRI dataset, and 0.74 for the ZNSDB dataset. 
\end{abstract}


\section{Introduction}
Image segmentation is the process of partitioning the pixels of an image into segments associated with different classes \cite{r13}. The main goal of segmentation is to simplify the representation of an image for analysis and interpretation of the contents present in the image. Image segmentation has been used in various computer vision tasks, such as object localization, boundary detection, and recognition. In essence, these tasks are performed by assigning a certain label to each pixel in the image. 

In the medical image analysis domain, segmentation holds fundamental importance for computer-aided diagnosis (CAD) systems. 
The segmentation extracts the region of interest (ROI) through a semi-automatic or automatic process. It divides an image into areas based on a specified description, such as segmentation of tissues in a brain MRI \cite{cnnMRI2016} or segmentation of lungs within a chest X-ray \cite{faizan2020}. 

Recently deep learning (DL) methods gained much attention due to their better performance for medical image segmentation tasks \cite{deeplearning2018, migan2018, faizan2020}. A deep learning neural network consists of a network comprising of multiple non-linear processing layers. 
Among deep learning methods, the encoder-decoder neural networks (such as the U-Net \cite{r4}) showed promising results on medical image segmentation. Since the inception of the U-Net model, many variants of the U-Net have been proposed by the deep learning research community. Few well known variants are UNET++ \cite{r14}, 3D U-Net, and Residual U-Net \cite{r16} and R2U++ \cite{mehreen2021}.

The aforementioned deep learning architectures have shown great potential on medical segmentation task. However, all of these algorithms need very high computational power, hardware resources and memory. Typically, these algorithms comprise of parameters in millions. For example, U-Net++ has 9.04 millions trainable parameters, Residual-Net18 has about 11 million parameters, and Residual-Net34 has 63.5 millions. These models require specialized computing hardware such as graphical processing units (GPUs) and Tensor Processing Units (TPUs). 
Hence, these algorithms are computationally expensive and their porting out to prototyping on portable devices is challenging. 
Therefore, to make medical image segmentation models portable and compatible for prototyping, we need to implement these models on low power egde devices. 
Enabling edge devices to perform medical image segmentation to aid in diagnosis can truly transform the healthcare ecosystem. Promising aspects include faster diagnosis, accessibility of AI-based analysis in remote areas, and portability of diagnosis devices. In addition, the portability of the devices will be supported by low-power requirements and reduced model size.
One approach to reducing the model sizes is using neural network compression techniques such as knowledge distillation or structural sparsity  \cite{hinton2015distilling, wen2016learning}. The working principle is based on the underlying assumption that non-critical parts of the networks can be reduced. However, for the implementation of medical image segmentation on the low-cost neural compute stick (NCS-2), we were interested in adopting a more straightforward path. Hence, we introduced modifications to the U-net model, and neural network compression methods are not directly covered in this work. We describe a modified, watered-down implementation of the U-Net segmentation model that is compatible with the NCS-2 resources. The proposed architecture implemented on NCS-2 can perform the segmentation task with high accuracy, comparable to the U-Net model. 
Our main contributions are given below:

\begin{itemize}
\item We present an optimized implementation of the U-Net model for medical image segmentation. Compared to the Intel's implementation with 7.85 million parameters \cite{intelunet}, the number of parameters is reduced to 0.49 million.
\item We present an implementation of the modified U-Net model on Neural Compute Stick 2 (NCS-2). 
\item We report results on three dataset and demonstrate that the model is able to perform competitively. 
\end{itemize}

\section{Background}
\label{sec:background}
\subsection{U-Net Model}
\label{sec:unetmodel}

Deep learning \cite{deeplearnigNATURE, schmidhuber2015deep} has greatly revolutionized many domains involving analysis of large images, audio, text, video, or tabular data. A major challenge that initially hampered the success of convolutional neural networks in the domain of medical Image segmentation was the unavailability of sufficient medical images for training deep learning models. To overcome this problem, a segmentation network called U-Net was proposed by \cite{r4}, which was specifically designed for the medical image segmentation task of smaller datasets.
The output of U-Net is a pixel-wise labeled image with a segmented region of interest. To perform the segmentation, the U-Net captures two types of information: What and Where. A typical convolutional neural network for image classification does not retain the 'where' information which is vital for segmentation tasks. Therefore, in order to get back the lost information, a corresponding up-sampling network is added in the U-Net architecture \cite{r4}. Thus, a U-Net comprises an encoder-decoder module.  

\textbf{Encoder:} 
The encoder is used to find out the context of an input image. The encoder identifies what is present in the input image. The encoder network has convolutional layers followed by max-pooling layers. The convolutional kernel is of size 3x3. In the first layer of the encoder, 64 filters are used. The number is continuously doubled in the subsequent layers of the encoder. 
The output from the encoder block is then down-sampled by the factor of 2 using max-pooling with the stride value set to 2. It reduces the dimension of the image by a factor of two. The process is repeated again in the succeeding encoder layers. As the image passes from each encoder layer, the size of the image will be reduced. On the other hand, the depth of the feature map will keep on increasing till the end of the encoder part \cite{r4}.

\textbf{Decoder:} 
The decoder is a symmetric expanding path for the encoder network. The decoder identifies the location of the region of interest in the input image. The decoder up-samples the output from the encoder via transposed convolution.
The output generated from upsampling is concatenated with the corresponding feature maps from the encoder network. It is followed by two consecutive convolution layers and the process is repeated till the end of decoder network. As the image passes from decoder layers, the feature maps keep on decreasing with increasing image resolution till we get the full size image.

The U-Net architecture was adopted quickly due to several advantages. Firstly, it captures context and location information simultaneously. Secondly, it gave state-of-the-art performance on segmentation task even for small medical imaging datasets. Finally, it is trained as an end-to-end model and provides segmentation mask in a forward pass. Consequently, the popularity of U-Net is not limited to medical imaging applications only and its effectiveness is also proven for other computer vision applications such as road segmentation \cite{recurrentunet2019}, \cite{r5}.

\subsection{Edge Computing}
\label{sec:edgecomputing}
Traditionally, as the requirements of the deep learning architecture would grow, one would need more computational resources and specialized hardware resources such as Graphics Processing Units (GPUs) and Tensor Processing Units (TPUs). 
Recently, NCS has gained attention as a cost-effective device for neural networks inference \cite{ncs}. NCS is mainly used at the end node of the network (edge computing) as it facilitates the computation and accelerates the machine learning model’s inference over edge devices \cite{ncs}. 
The NCS device is very helpful in development of offline artificial intelligence prototyping as it provides support for compiling, tuning, validating and accelerating the models. It also supports libraries like Caffe and TensorFlow, in addition to the self-designed models \cite{r6}. 

\subsection{Recent work on NCS for deep learning inference}
NCS has recently been used for inference of deep learning models for computer vision applications. Xu et al., \cite{classifyvoxel2017} demonstrated the generation of synthetic 3D point-clouds using a 3D CNN. They used 3D point cloud occluded volumes depicting real world scenes for training the network and then compared the computational time and required power for running an inference task across multiple devices like the Nvidia Titan X, an older version of the NCS (Fathom NCS) and Intel i7-5930K CPU. Through experimental evaluations in terms of inference, time (milliseconds), and power (watts), they demonstrated that the Fathom NCS performed significantly better and provided a low-cost and low-power solution for inference purposes. The key observation from \cite{classifyvoxel2017} is that the Fathom NCS used the least power and completed the inference task at hand reasonably fast without any loss of accuracy that was obtained through other conventional methods.

Naman et al., \cite{r22} optimized a model for classification of book cover images into genres and then compiled the model for inference on an NCS. Xing et al., \cite{r23} proposed a CNN with Raspberry Pi based system that runs a pre-trained inference model using NCS. The model performed facial emotion recognition using images. Othman et al., \cite{r24}, presented a method for real time object detection and recognition that runs at high frames per second (FPS) on an embedded platform using Movidius NCS.

To the best of our knowledge, the use of NCS-2 inference for medical image segmentation is not explored yet. In our work, we explore the potential of NCS-2 on medical image segmentation tasks and present a modified U-Net model that provides compatibility with the NCS-2 resources.

\section{Datasets}
\label{sec:datasets}
We used three datasets including two subsets of the medical segmentation decathlon dataset (from now onwards, we call it the decathlon dataset) \cite{decathlon2021}. The third dataset comprises Ziehl–Neelsen sputum smear microscopy images acquired from publicly available dataset \cite{r9}. We briefly describe the three datasets in the following text.

\subsection {Medical Segmentation Decathlon dataset}
\begin{figure}[!ht]
\centering
\includegraphics[width=0.9\linewidth]{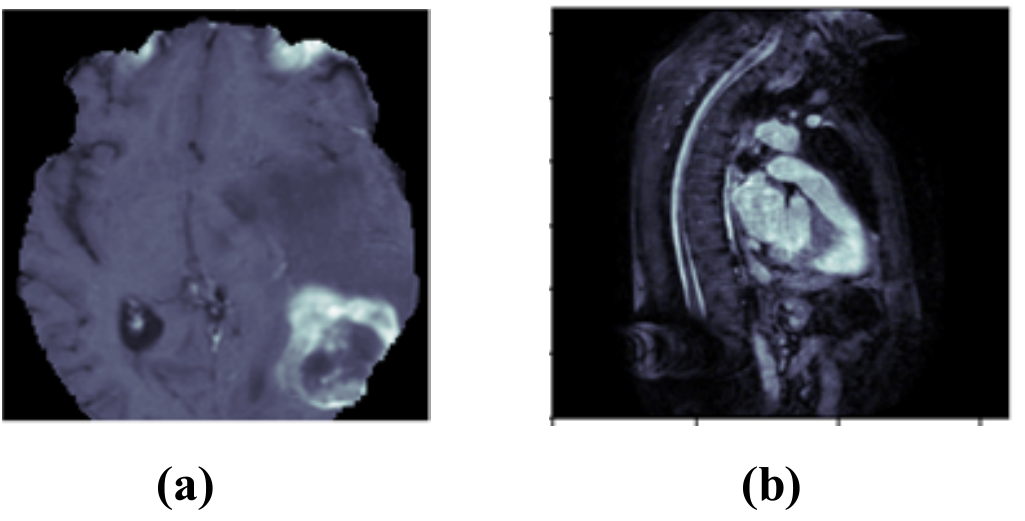}
\caption{Sample images from the decathlon dataset. (a) Brain MRI from BraTs data. (b) Heart MRI image.}
\label{fig:sampledecathlon}
\end{figure}

This dataset is large collection of annotated medical image datasets of various clinically relevant anatomies \cite{decathlon2021}. The images are available under open source license to facilitate the development of semantic segmentation algorithms. The dataset is a result of contributions of images from multiple institutions. 
The images are  collected across multiple anatomies of interest and multiple modalities. 
In our experiments, we have used two subsets of the decathlon dataset, (i) a subset of the BraTs dataset which contains 75020 images of size 128 $\times$ 128 pixels, and (ii) Heart MRI images dataset containing 2271 images of size 340 $\times$ 340 pixels. Sample images from the decathlon dataset are shown in Figure \ref{fig:sampledecathlon}.


\subsection{Ziehl-Neelsen Microscopy Images Dataset}
Detection of Tuberculosis (TB) bacteria in sputum smear microscopy is typically performed through manual processes, which requires expert pathologists. The diagnosis process can be automated by using deep learning (U-Net) model for segmentation of the bacilli. In our experiment, we acquire the Ziehl-Nelseen stained sputum smear microscopy images from publicly available dataset i.e. ZNSDB \cite{r9}. The dataset consists of various categories i.e. overlapping bacilli, autofocused data, occluded bacilli, over-stained views with bacilli and artifacts, single or few bacilli, images without bacilli. 
We used 570 images selected from all the dataset categories. Out of these, 85$\%$ of data is selected at random for training and validation and 15$\%$ is used for testing the model.
\begin{figure}[!ht]
\centering
\includegraphics[width=\linewidth]{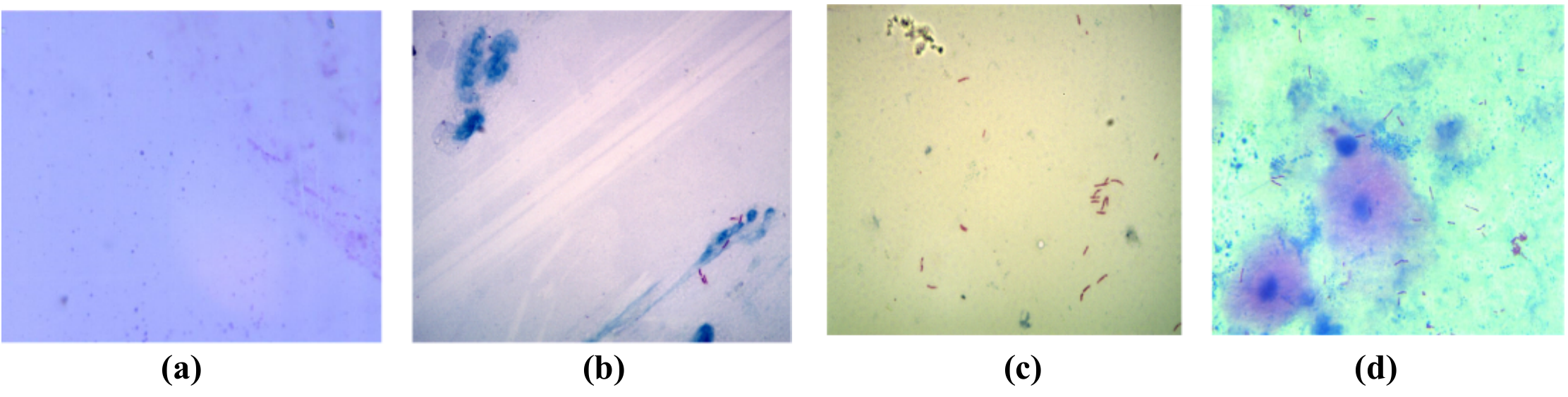}
\caption{Sample images of different categories from Ziehl - Neelsen sputum microscopy images dataset. (a) No bacilli. (b) Very few bacilli. (c) Overlapping bacilli. (d) Bacilli and artifacts.}
\label{fig:znsamples}
\end{figure}
\section{Methodology}
\label{sec:methodology}
In this work, we used the above mentioned three datasets to evaluate our proposed model. The methodology we followed is briefed in this section. 
To develop deep learning inference application on edge device, we used energy-efficient and low-cost Movidius NCS-2. The NCS provides profiling, tuning, and compiling a deep neural network on a development computer using OpenVINO development kit \cite{openvino}. 

\subsection{Evaluation of Intel's U-net model}
The Intel research team presented a basic U-Net model for medical image segmentation \cite{ncs, intelunet}. They trained and evaluated the model on BraTs dataset that contains brain MRI images \cite{decathlon2021} and performed inference on NCS-2.
The distribution of training and test sets is given in Table \ref{tab:distributionofdata}. 

\begin{table}[!th]
\caption{Dataset description used in Intel's experiment}
\label{tab:distributionofdata}
\centering
\begin{tabular}{p{2.2in}r}
\hline
Dataset       & BraTs       \\\hline
Image Size    & 128x128   \\\hline
Training Images & 59985   \\\hline
Validation Images       & 5890  \\\hline
Test Images   & 9145   \\          
\hline
\end{tabular}
\end{table}

\begin{figure}[!th]
\centering
\includegraphics[width=0.8\columnwidth]{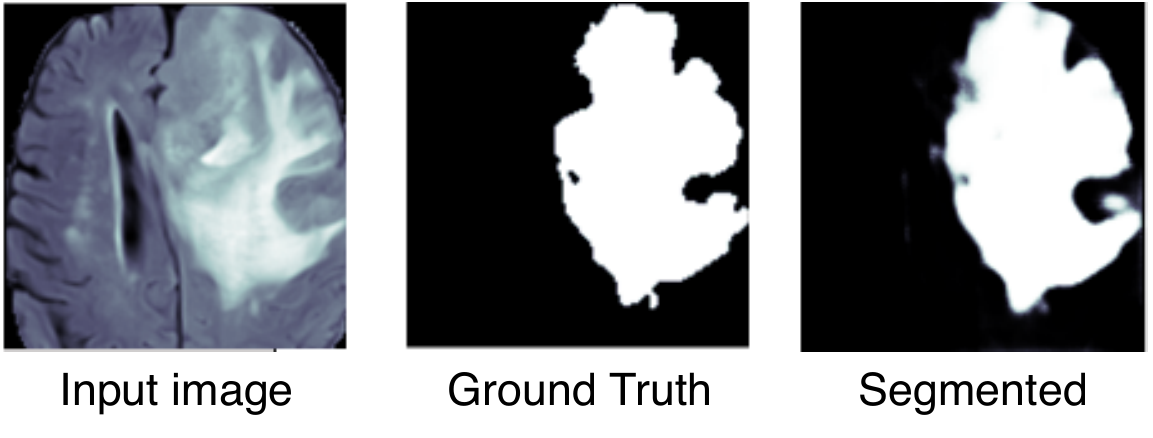}
\caption{\textbf{Results for Intel's U-Net}. Sample image is shown for brain MRI from BraTs dataset with corresponding ground truth segmented mask and output segmented image. }
\label{fig:intelresult}
\end{figure}

To compare with the Intel's model, we performed training and testing using the same configurations on the three datasets \cite{intelunet}. Sample of the segmentation results for the Intel's UNet model is shown in Figure \ref{fig:intelresult}. The model configuration and the results in terms of dice score are summarized in Table \ref{tab:comparison}.

\subsection{Modification of Intel's U-Net Model}
\begin{figure}[!h]
\begin{centering}
\includegraphics[width=\columnwidth]{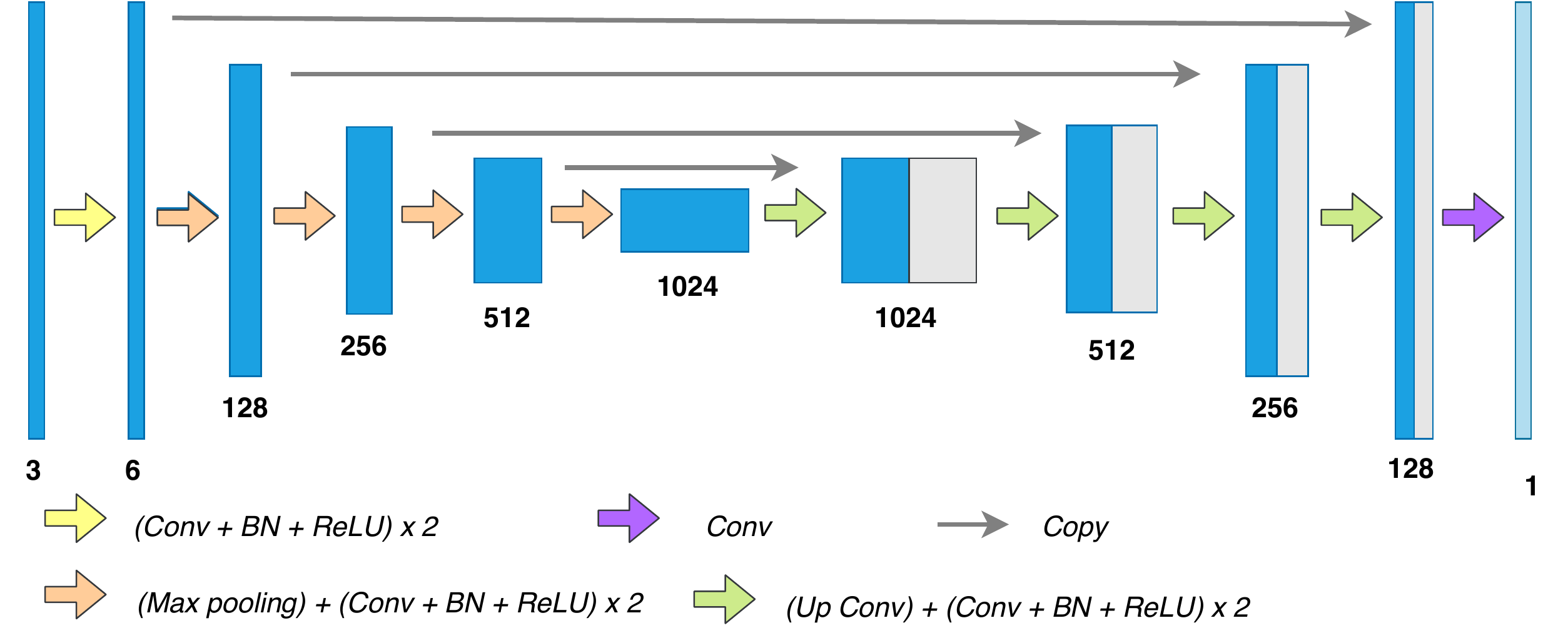}
\caption{Proposed Model Architecture (A modified U-Net). The number below each block shows the number of channels. The different operations are denoted by the arrows. Conv: Convolution. BN: Batch Normalization.}
\label{fig:proposedarchitecture}
\end{centering}
\end{figure}

In our proposed model, we presented a modified form of the basic U-Net model by customizing the layers to optimize the model for inference on NCS-2. For example, we reduced the number of features maps, which in turn help to reduce the number of parameters and the required resources. We also added batch normalization layers and maxpooling layers in the basic model. The modified architecture of proposed model is shown in Figure \ref{fig:proposedarchitecture}.

The purpose of the modifications is to optimize the selected model in terms of resources, features and parameters to make it compatible for NCS-2. The original U-Net model contains 30 million parameters (with 32 feature maps) exceeding the memory capacity of NCS-2. In comparison, we reduced the parameters to 0.49 million (see Table \ref{tab:comparison}).

We performed training on Intel Core (TM) i7-6700 CPU 3.40GHz equipped with an NVIDIA GeForce RTX 2060 with 6 GB memory.

\begin{table}[!h]
\caption{Results comparison between basic U-Net model, Intel's model and the proposed modified U-Net model using BraTs dataset}
\label{tab:comparison}
\begin{center}
\begin{tabular}{lrrr}
\hline
Model     & Basic U-Net & Intel's Model & Proposed Model      \\\hline
Epochs   & 100 & 30 & 30  \\\hline
Parameters & 30 million & 7.85 million & 0.49 million \\\hline
Feature Maps & 64 & 32  & 8 \\\hline
Batch Norm Layer      & Nill & Nill & included \\\hline
Highest Dice Score  & 0.97 & 0.9625 & 0.9625 \\   \hline
Test time on NCS   & N/A$^*$ & 123 msec  & 61 msec \\
\hline
\multicolumn{4}{l}{$^*$The model cannot be ported to NCS-2.}
\end{tabular}
\end{center}
\end{table}

\begin{figure}[h!]
\centering
\includegraphics[width=0.85\columnwidth]{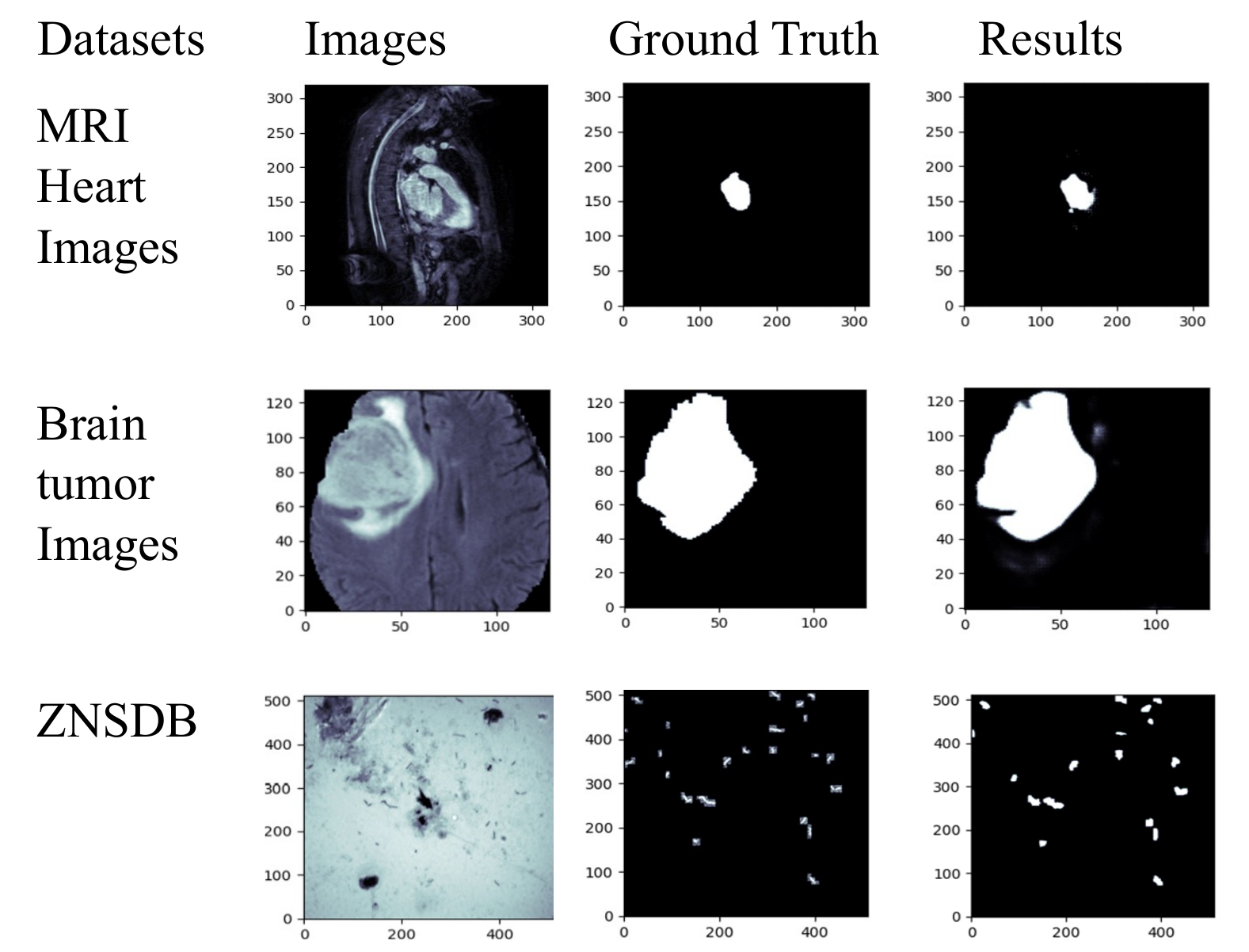}
\caption{Segmentation results for sample images taken from heart MRI images, BraTs and ZNDB datasets using proposed model. A closed match can be seen for the segmented mask in the output images compared with the ground truth that demonstrates the effectiveness of the proposed architecture.} 
\label{fig:allresults}
\end{figure}

\section{Results and Discussion}
\label{sec:results}

\subsection{Comparison for BraTs (brain MRI) dataset}
For comparison purposes, we evaluated our model on the BraTs dataset as this was used by Intel's model \cite{intelunet}. With our modified model, we were able to achieve an optimized architecture with comparable results. The comparison is made in terms of \emph{feature maps, resources, number of parameters, and highest dice score.} 
We set the learning rate to $1e^{-5}$, and run the training for 30 epochs. The optimal values for the learning rate, number of epochs, input image sizes, and feature maps were chosen after multiple runs of the experiment.  The configuration and the dice scores for the different datasets are reported in Table \ref{tab:results}.
Figure \ref{fig:allresults} shows samples of results (the input image, the ground truth and the output segmented mask) obtained by the proposed method for each dataset. 
\begin{table}[H]
\caption{Results of the proposed model on the different datasets with configuration and datasets' description.}
\label{tab:results}
\begin{center}
\begin{tabular}{lrrr}
\hline
Dataset       & BraTs & Heart MRI Images & ZNSDB Images
      \\\hline
Image Size    & 128x128 & 340x340 & 512x512  \\\hline
Training Images & 59985 & 1819 & 397  \\\hline
Validation Images       & 5890 & 222 & 36 \\\hline
Testing Images   & 9145 & 230 & 37  \\\hline   
Epochs   & 30 & 50 & 50  \\\hline
Avg. test time & 61 ms
 & 324 ms
 & 236 ms
 \\\hline
Feature Maps & 8 & 8  & 8 \\\hline
Batch Norm Layer      & included & included & included \\\hline
Highest Dice Score  & 0.9625 & 0.94 & 0.74 \\  
\hline
\multicolumn{4}{l}{Note: Avg test time = Average time per test image. ms = millisecond.}\\
\end{tabular}
\end{center}
\end{table}

\subsection{Discussion}
\begin{figure}[!th]

\includegraphics[width=0.9\columnwidth]{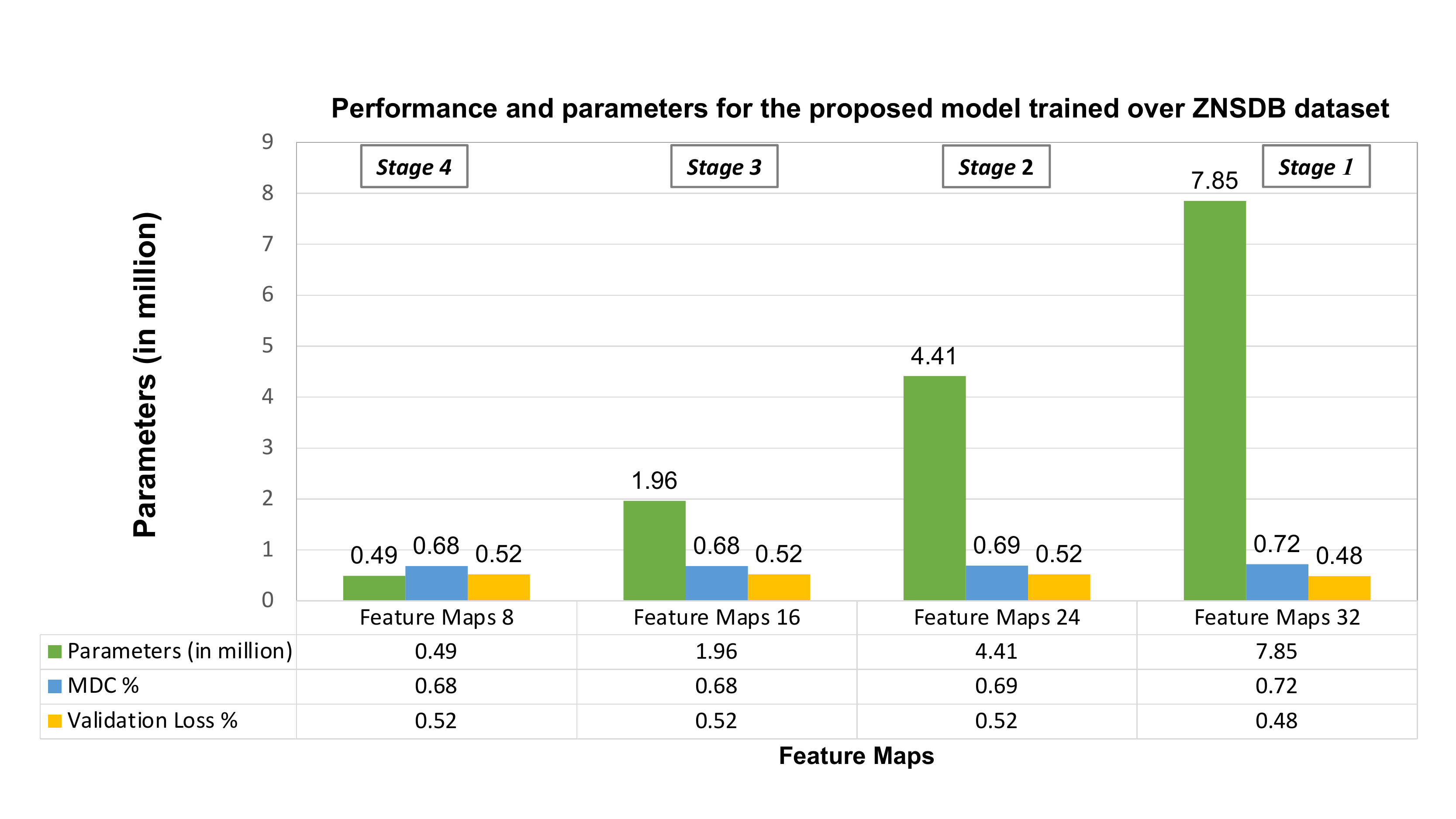}
\caption{Performance and parameters of the proposed model - shown with respect to increasing the number of features maps. The trends are shown for the Ziehl-Nelseen dataset. For a 94\% reduction in the number of parameters from Stage 1 to Stage 4, the drop in dice score is only 4\%. MDC: Maximum Dice Coefficent}
\label{fig:performancechart}
\end{figure}

We concluded that using our proposed model; we can perform segmentation tasks of medical images like MRI or microscopy images and implement our model over NCS-2 as the proposed model requires fewer resources. In addition, our proposed optimized segmentation model has performed segmentation with the highest dice score of greater than 90 percent for each dataset using NCS-2. Figure \ref{fig:performancechart} shows the resources and performance chart of the proposed model for Ziehl - Nelseen microscopy images dataset. Obviously, when we reduce the number of parameters, the performance degradation in terms of the validation loss and the maximum dice score is nominal. For example, from stage 1 through stage 4, the number of parameters is reduced from 7.85 million to 0.49 million (a 94\% reduction), resulting in a drop of dice score from 0.72 to 0.68 (a 4\% drop only). This is a trade-off that we make to achieve a significant reduction in the model size. 

\subsection{Potential Applications and Limitations}
We believe that the proposed implementation of medical image segmentation on NCS-2 will enable the design of cost-effective and portable solutions for the medical imaging and diagnostics industry. The potential target users are hospitals, diagnostics laboratories, and research laboratories.

In this work, we have used pre-processed images from the available datasets. But in clinical applications, medical images may come with noise at the human or machine level and require pre-processing before using them for segmentation or classification. The model inference presented in this work is not validated by clinicians, so the use of the model should be done with precaution. 
We advise that interpretation should be made with care, and input of the trained professionals and medical imaging domain experts should be sought where needed.
We aimed to minimize the resource requirements while achieving acceptable performance and the performance as reported in this work may not be the best compared to the performance achieved on high-end computational devices. So, a further improvement in the accuracy is also desired to improve the overall performance of the proposed architecture. 
In the case of Ziehl–Neelsen dataset, the experimental results also show that performance degrades for images where the bacilli are out of focus, or the images contain artifacts such as irregular staining.
NCS-2 hardware has no cooling package. Hence, the operation may pose challenges in higher temperature environments. In addition, the working is limited by the lack of embedded devices for NCS-2 operation though the manufacturer in future design may address this.

\section{Conclusion}
\label{sec:conclusion}
In this work, we presented an optimized implementation of the U-Net model to perform the segmentation of medical images on the NCS-2. We introduced modifications in the U-Net architecture to reduce the number of parameters and optimize the model for the required computational resources. Through experimental results, we have demonstrated that the proposed implementation works fine with a significantly lower number of parameters. The performance of the modified model is evaluated for dice score and compared in terms of the number of parameters, feature maps. The model is tested on three medical imaging datasets: BraTs, heart MRI, and Ziehl–Neelsen microscopy images datasets.
The proposed model provided inference on NCS-2 and achieved segmentation task with the highest dice scores of 0.9625 for the BraTs dataset, 0.74 on the ZNSDB dataset, and 0.94 on the heart MRI dataset. 
We believe that this work will advance the research on medical image diagnosis using deep learning models on low-cost hardware.



\newpage



\vspace{11pt}


\vfill

\end{document}